\documentclass[fleqn,usenatbib]{mnras}

\usepackage[T1]{fontenc}

\DeclareRobustCommand{\VAN}[3]{#2}
\let\VANthebibliography\thebibliography
\def\thebibliography{\DeclareRobustCommand{\VAN}[3]{##3}\VANthebibliography}

\usepackage{graphicx}	% Including figure files
\usepackage{amsmath}	% Advanced maths commands
\usepackage{amssymb}	% Extra maths symbols

\newcommand{\target}{MACS J1149.5+2223}
\newcommand{\wfcir}{WFC3/IR}
\newcommand{\hst}{{\it HST}}
\newcommand{\hubble}{{\it Hubble}}
\newcommand{\euclid}{{\it Euclid}}
\newcommand{\romantel}{{\it Roman}}
\newcommand{\rtwo}{$r_{200m}$}
\newcommand{\mtwo}{$M_{200m}$}

\title[ICL in \target]{Discovery of a Possible Splashback Feature in the Intracluster Light of \target}

\author[A.~H. Gonzalez et al.]{
Anthony H. Gonzalez,$^{1}$\thanks{E-mail: anthonyhg@ufl.edu (AHG)}
Tyler George,$^{1}$
Thomas Connor,$^{2,3,4}$
Alis Deason,$^{5,6}$
\newauthor
Megan Donahue,$^{7}$
Mireia Montes,$^{8}$
Ann I. Zabludoff,$^{9}$
and Dennis Zaritsky$^{9}$
\\
$^{1}$Department of Astronomy, University of Florida, Gainesville, FL 32611, USA\\
$^{2}$Jet Propulsion Laboratory, California Institute of Technology, 4800 Oak Grove Drive, Pasadena, CA 91109, USA\\
$^{3}$The Observatories of the Carnegie Institution for Science, 813 Santa Barbara St, Pasadena, CA 91101, USA\\
$^{4}$NPP Fellow\\
$^{5}$Institute for Computational Cosmology, Department of Physics, University of Durham, South Road, Durham DH1 3LE, UK\\
$^{6}$Centre for Extragalactic Astronomy, Department of Physics, University of Durham, South Road, Durham DH1 3LE, UK\\
$^{7}$Department of Physics and Astronomy, Michigan State University, East Lansing, MI 48824, USA\\
$^{8}$Space Telescope Science Institute, Baltimore, MD 21218, USA
\\
$^{9}$Department of Astronomy \& Steward Observatory, University of Arizona, Tucson, AZ 85721, USA\\
}

\pubyear{2021}

\begin{document}
\label{firstpage}
\pagerange{\pageref{firstpage}--\pageref{lastpage}}
\maketitle

\begin{abstract}
We present an analysis of the intracluster light in the Frontier Field Cluster \target\ ($z=0.544$), which combines new and archival \hubble\ \wfcir\ imaging  
to provide continuous radial coverage out to 2.8 Mpc from the brightest cluster galaxy. Employing careful treatment of potential systematic biases and using data at the largest radii to determine the background sky level, we reconstruct the surface brightness profile out to a radius of 2 Mpc. This radius is the largest
to which the intracluster light (ICL) has been measured for an individual cluster. Within this radius, we measure a total luminosity of $1.5\times10^{13}$ L$_\odot$ for the brightest cluster galaxy plus ICL light.  From the profile and its  logarithmic slope, we identify
the transition from the brightest cluster galaxy to ICL at $r\sim70$ kpc. Remarkably, we also detect an inflection in the profile centered in the $1.2-1.7$ Mpc ($0.37-0.52$ \rtwo) radial bin, 
a signature of an infall caustic in the stellar distribution. 
Based upon the shape and strength of the feature, we interpret it 
as potentially being at the splashback radius,   
although the radius is smaller than theoretical predictions. If this is the splashback radius, then it is the first such detection in the ICL and the first detection of the splashback radius for an individual cluster.  
Similar analyses should be possible with the other Frontier Field clusters, and eventually with clusters 
from the \euclid\ and \romantel\ missions.
\end{abstract}

\begin{keywords}
galaxies: clusters: general -- galaxies: evolution
\end{keywords}

%%%%%%%%%%%%%%%%%%%%%%%%%%%%%%%%%%%%%%%%%%%%%%%%%%

\section{Introduction}

 It has been seven decades since \citet{Zwicky1951} first noted
 extended, diffuse emission in the Coma cluster. 
  Because this diffuse emission is typically centered on the brightest cluster galaxy \citep[BGC; e.g., ][]{schombert1988,Gonzalez2005}, in early literature this component was often referred to as a cD envelope \citep[e.g., ][]{oemler1973}, but is now most commonly referred to as intracluster light (ICL). The stars that comprise the ICL are by definition unbound from individual galaxies, and instead orbit in the gravitational potential well of the cluster. 
ICL studies have gained in prominence in recent years, driven primarily by the realization that this component encodes key information about the global cluster properties and the evolution of galaxy clusters \citep[e.g.,][]{montes2019b}.

Specifically, there are three dominant questions that have motivated research on the ICL. 
The first is how much mass the ICL contributes to the baryon budget in clusters. A number of authors find that the ICL contains a non-negligible fraction of the total stellar mass in cluster cores
\citep[e.g.,][]{Gonzalez2007,Gonzalez2013a,Lagana2013,Furnell2021}, and thus cannot be ignored in a census of cluster baryons. However, until the recent work of \citet{sampiaosantos2020}, even the studies with the greatest radial coverage \citep{Zibetti2005, zhang2019} lacked the combination of radial range and surface brightness sensitivity to reach a point where the measured total stellar mass in the ICL converges, and only one published result for an individual cluster reached a projected radius of 1 Mpc \citep[Abell 1413,][]{schombert1986, schombert1988}.

The second open question is the origin and assembly history of the ICL, which is central to understanding the global evolution of the cluster galaxy population. Recent observational studies find significant late-time build-up of the ICL \citep{DeMaio2020,Furnell2021} and inside-out growth  \citep{DeMaio2020}. Whether the bulk of the ICL arises from tidal disruption of low mass dwarf galaxies \citep{Giallongo2014,Annunziatella2016} or tidal stripping of the outskirts of more massive galaxies \citep{Montes2014,DeMaio2015,montes2018,DeMaio2018} remains an unsettled question, 
impeding interpretation of 
integrated cluster galaxy properties such as the luminosity function.

The third question centers on how well the ICL traces the gravitational potential of the host cluster. \citet{montes2019} demonstrated for the \hubble\ Frontier Field clusters that the ICL is an accurate luminous tracer of the lensing mass distribution within the central 140 kpc, and hence  argued that the ICL traces the dark matter significantly better than X-ray emission. Subsequent investigation by \citet{alonsoasensio2020} using the C-EAGLE simulations indicates that the intrinsic correlation is even tighter than observed by \citet{montes2019}. As such, the ICL potentially provides an efficient means of mapping out cluster mass distributions in future surveys with the \euclid\ and \romantel\ missions.
\citet{sampiaosantos2020} have also recently argued that the total ICL stellar mass scales with cluster mass, providing a low-scatter proxy for total cluster mass and further enhancing the potential utility of ICL for future cluster studies.

Beyond the questions above, there is another area in which the ICL may be of particular use: defining the halo boundary and connecting observations with simulations. 
When matter falls into a halo, the apocenter of its initial orbit constitutes an observable physical radius 
\citep{gunn1972,fillmore1984,bertschinger1985}.  \citet{diemer2014} pointed out that  this ''splashback'' radius  
is often characterized by a sharp break in the density profile.
The location of this feature can be directly predicted by simulations; 
at a given redshift the radial location of this feature is determined by cluster mass and accretion rate \citep{diemer2014,Adhikari2014,More2015,Diemer2020b,Oniel2020,Xhakaj2020}. 

Recent observational programs have detected the splashback radius for ensembles of stacked clusters via galaxy surface density profiles \citep{more2016,Zucker2019,shin2019,Adhikari2020,murata2020,bianconi2020} or  weak gravitational lensing \citep{Contigiani2019}, confirming the existence of this feature. Of direct relevance to the present work, \citet{deason2021} used the C-EAGLE simulations to demonstrate that this signature is expected in the intracluster light.  
Given the results of \citet{sampiaosantos2020}, if the total ICL stellar mass and splashback radius can be measured, then 
in principle one can determine both  accretion histories and cluster masses.
Moreover, measurements of the splashback radii for individual clusters would facilitate environmental studies of the cluster galaxy population, particularly enabling a clear identification of currently infalling galaxies beyond the splashback radius.

Here we focus on the cluster \target\ \citep[$z=0.544$,][]{ebeling2001,ebeling07}. This cluster was part of the Cluster Lensing and Supernova survey with \hubble\ \citep[CLASH]{Postman2012} and is one of the six \hubble\ Frontier Field clusters \citep[HFF,][]{lotz2017}. 
In addition to being an extremely massive cluster \citep[$M_{200m}=3.4\pm0.7\times10^{15}$ M$_\odot$;][]{umetsu2014}, it is known to be a complex ongoing merger.
\citet{golovich2016} find evidence for three distinct merging components, including a major merger of roughly equal mass components plus accretion of a third, lower mass subcluster.  These properties and the extensive existing data make \target\ a unique system for a detailed investigation of the distribution and properties of the ICL, and these data have been previously used by multiple groups to investigate the ICL in the cluster core \citep{Morishita2016,montes2018}. In this program we significantly extend the radial regime over which the ICL can be studied.

Using a combination of new and archival \hst\ observations, we trace the surface brightness profile of the ICL out to 2 Mpc. We quantify the total luminosity contained within this diffuse component and demonstrate the feasibility of probing the ICL to large radii with upcoming space-based survey missions.  We also investigate whether we can detect caustic signatures -- edges in the matter phase space distribution that are detectable as edges in the density distribution --  in the radial profile, as predicted by \citet{deason2021}. We save consideration of the two-dimensional distribution of the ICL, and of the origin of the ICL, for future companion papers.
Throughout this paper \rtwo\ is defined relative to the mean density of the Universe, and we use the cosmological parameters of the \citet{planck2015}. Specifically, $H_0=67.7$ km s$^{-1}$ Mpc$^{-1}$, $\Omega_m=0.307$, and $\Omega_\Lambda=0.693$.

\section{\hst\ Data}

The data used in this analysis consist of a combination of all archival F105W and F160W data for \target\ taken prior to 2019, including associated parallel field data,  and new observations from program HST-GO-15308 (PI: Gonzalez) in these same filters. The choice of filters is based upon similar criteria to those used for \citet{DeMaio2018}, as explained below. We focus our attention on \wfcir\ data due to its sensitivity to emission from relatively old stellar populations, for which the spectral energy distributions peak in the near-infrared. The predominant source of archival data for this program is the HFF program, which observed \target\ in F105W, F125W, F140W, and F160W. The F105W and F160W filters together provide the longest wavelength lever arm, which is beneficial for identifying color gradients in the ICL. A consideration with F105W data is that there exists a time-variable excess Earth glow in this passband due to the Helium 1.038 $\mu$m emission line (see Instrument Science Report WFC3 2014-03). To minimize the impact of this Helium line, we acquired F105W observations during the middle of each orbit when the Earth glow would be lowest. 

The program IDs, dates, number of exposures, total exposure times, and filters for all archival data used in this analysis are listed in Table \ref{tab:data}. The column {\it location} indicates whether the pointing was centered on the cluster core (CORE) or a parallel field (PAR). 

\begin{figure*}
\includegraphics{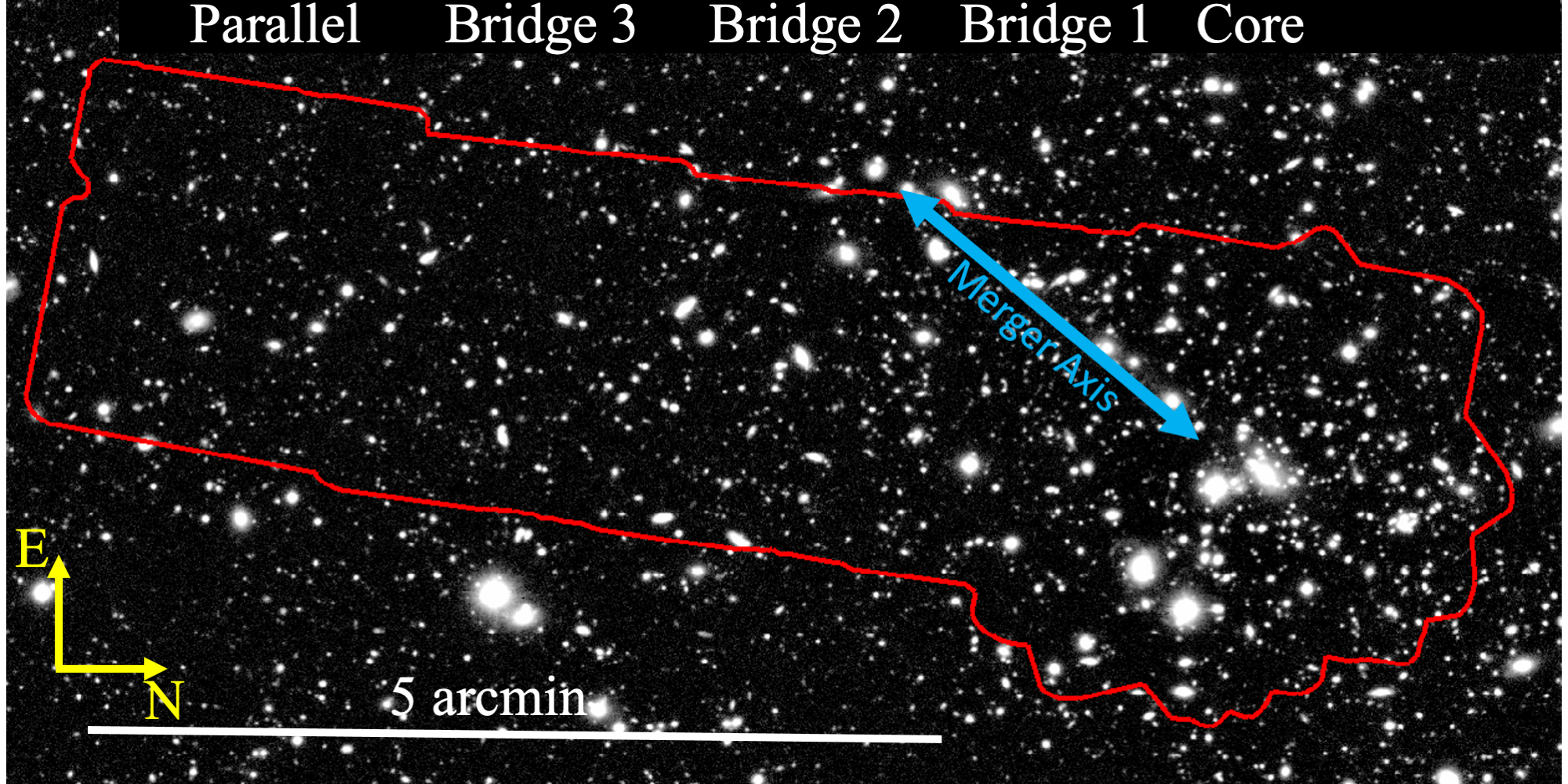}
\caption{Shown here is the coverage of the HST \wfcir\ F105W data (red contours) overlaid on a Subaru $z-$band image from CLASH \citep{Postman2012}, illustrating the geometry of the bridge fields relative to the core and parallel fields. North is to the right, and east is up. The core and parallel fields are primarily from the CLASH \citep{Postman2012} and  \hubble\ Frontier Field  \citep[HFF;][]{lotz2017} surveys. Each bridge field overlaps by approximately 20\% with adjacent fields. Together these fields enable continuous radial coverage and correction for sky level offsets between the core and parallel observations. The blue arrow illustrates the axis of the ongoing cluster merger \citep{golovich2016}, which is oriented approximately 30 degrees from the long axis of our \hst\ imaging. For scale, 5$^\prime$ corresponds to 1.97 Mpc at the cluster redshift.}
\label{fig:geometry}
\end{figure*}

The new observations consist of three bridging fields designed to uniformly span the gap between the cluster core and parallel fields. The geometry of these fields is schematically shown in Figure \ref{fig:geometry}. We refer to these fields as BRIDGE1, BRIDGE2, and BRIDGE3 in Table \ref{tab:data}, where we present the exposure information.  The exposure times for the bridge fields correspond to 1 orbit per filter per pointing. These modest observations are sufficient because we are limited in this analysis by systematic rather than statistical uncertainties at large radii.

\begin{table}
	\centering
	\caption{Data sets included in this analysis.}
	\label{tab:data}
	\begin{tabular}{lccrr} % four columns, alignment for each
		\hline
		PID & Filter  & Dates & \multicolumn{1}{c}{\# of} & \multicolumn{1}{c}{Time} \\ %Time (ks)  \\
		& & & \multicolumn{1}{c}{Images} & \multicolumn{1}{c}{ (ks)}\\
		\hline
		\multicolumn{5}{c}{CORE}\\
		\hline
		12068 & F105W & 30 Jan 2011 - 27 Feb 2011 & 5 & 2.8 \\
		     & F160W & 04 Dec 2010 - 09 Mar 2011 & 24 & 13.5 \\
		13504 & F105W & 21 Nov 2014 - 05 Jan 2015 & 48 & 67.3 \\
		     & F160W & 02 Nov 2013 - 05 Jan 2015 & 48 & 66.1 \\
	    13790 & F105W & 20 Apr 2015 - 13 Jul 2015 & 16 & 6.3 \\
		     & F160W & 29 Nov 2014 - 21 Jul 2015 & 63 & 24.0 \\
		14199 & F105W & 12 Feb 2016 - 14 Feb 2016 & 8 & 6.8 \\
		     & F160W & 30 Oct 2015 - 30 Oct 2016 & 71 & 29.6 \\
		14208 & F105W & 16 May 2016  & 3 & 1.2\\
		& F160W & 16 May 2016 & 4 & 2.3 \\
		14528 & F160W & 14 Jul 2016 - 20 Jul 2016  & 9 & 3.6\\
		14872& F160W & 05 Dec 2016   & 3 & 1.2\\
		\hline
		\multicolumn{5}{c}{PARALLEL}\\
		\hline
		13504 & F105W & 19 Apr 2015 - 09 May 2015 & 48 & 64.4 \\
		     & F160W & 14 Apr 2014 - 09 May 2015 & 48 & 66.5 \\
		\hline
		\multicolumn{5}{c}{BRIDGE1}\\
		\hline
		15308 & F105W & 25 May 2018 & 4 & 2.4 \\
		     & F160W & 25 May 2018 & 4 & 2.4 \\
		\hline
		\multicolumn{5}{c}{BRIDGE2}\\
		\hline
		15308 & F105W & 25 Apr 2018 & 4 & 2.4 \\
		& F160W & 25 Apr 2018 & 4 & 2.4 \\
		\hline
		\multicolumn{5}{c}{BRIDGE3}\\
		\hline
		15308 & F105W & 26 May 2018 & 4 & 2.4 \\
		& F160W & 26 May 2018 & 4 & 2.4 \\
		\hline\hline
	\end{tabular}
\end{table}

\section{Methods}
\subsection{Data Processing}

We reprocess the data starting with the calibrated individual exposures ({\tt flt} images) in a similar fashion to \citet{DeMaio2018}. All images are processed with the standard {\tt astrodrizzle} routines \citep{astrodrizzle2012_revised} and drizzled onto a common reference frame with a pixel scale of 0.1$^{\prime\prime}$. One important difference relative to the standard reduction provided by STScI is that we do not subtract the sky from the individual images to avoid unintentional subtraction of the ICL. As in \citet{DeMaio2018}, we also use data obtained at a similar epoch to each observation to apply a flatfielding correction to each image.  

The current processing does diverge from that of \citet{DeMaio2018} in several significant ways. First, in \citet{DeMaio2018} a planar gradient was subtracted from each exposure. We omit this step in the current analysis to fully preserve the ICL signal at large radii. Second, 
we mask pixels that are potentially contaminated by residual charge from previous observations (persistence). Third, we generate epoch-specific flatfields specifically for this analysis. Finally, there are some minor changes in the implementation of masking more generally to ensure consistency between overlapping images.  Details of the persistence masking, flatfield correction, and masking implementation are described below.

\subsubsection{Persistence}
\label{sec:persist}
Bright sources observed by the \wfcir\ detector leave residual charge after the observation which can result in an observed excess in the counts at that location in subsequent images \citep{smith_persistence_2008}. The level of this persistence is a function of time after the observation and the brightness of the previous source. Persistence maps for all \wfcir\ observations are provided by the WFC Persistence Project through the Mikulski Archive for Space Telescopes (MAST).\footnote{see \url{https://archive.stsci.edu/prepds/persist}} These maps include the impact of images taken up to 16 hours prior to the observations.  It was emphasized by \citet{borlaff2019} that images taken even further prior to an observation can still have a noticeable impact at the lowest surface brightness levels, 
and for their analysis of the \hubble\ Ultra Deep Field those authors generated persistence maps using data from the previous 96 hours. However, for the current analysis, in which we are focusing upon the radially averaged profile, the default persistence images are sufficient. 

In the pipeline, for individual exposures we mask all pixels for which the persistent charge exceeds a level equivalent to $\mu=33$ mag arcsec$^{-1}$. This is over a magnitude below the faintest level to which we trace the F105W surface brightness profile, and the impact of persistence is further diluted by the radial binning.  We also reject 18 F160W images of the cluster core (4\% of the total data, 3 from program 13790 and 15 from program 14199) that are severely compromised by persistence across much of the field.

\subsubsection{Flatfields}

As has been pointed out in several papers  \citep[][and references therein]{borlaff2019,DeMaio2020}, the sensitivity of the \wfcir\ detector has a long-term temporal component that for ICL analyses can be a dominant systematic uncertainty if data are processed using only the default \wfcir\ flats.
For this analysis, we therefore generate flatfield corrections (hereafter $\Delta$ flats) for a series of epochs corresponding to the \target\ observation epochs.

We start with all \wfcir\ science observations in the F105W and F160W filters from the period of 2010 Jan through 2019 Dec that were publicly available in the \hst\ archive as of March 2020.
From this sample, we next exclude observations that (i) have short exposure times of $<400$s in F105W or $<300$s in F160W, or (ii) target regions for which extended structures would be expected to compromise the flatfield (including extended low-redshift galaxies, galaxy clusters, regions with significant nebulosity, and dense stellar fields). All remaining images are then visually inspected by a member of our team (TG) to remove any additional images that are otherwise unsuitable for inclusion when generating a flat. This visual inspection removed approximate 2\% of the images.

For this restricted set of observations, we then define a set of fiducial dates and use an initial window of $\pm6$ months to generate lists of images from which to construct $\Delta$ flats. We further require a minimum of 250 (200) input images in F160w (F105W) for each $\Delta$ flat, and therefore widen time windows as necessary to reach this threshold. This is generally more of an issue for F105W, for which the final windows range from $\pm270$ days to $\pm420$ days.
The list of fiducial dates, window widths, and number of input images are given in Table \ref{tab:flats}.

\iffalse
\begin{table}
	\centering
	\caption{Old Definition of epochs for construction of $\Delta$ flats.}
	\label{tab:flats}
	\begin{tabular}{lccr} % four columns, alignment for each
		\hline
		Date & Half-Width  & \#  \\
	         & (days) &  Images \\
		\hline
		\multicolumn{3}{c}{F105W}\\
		\hline
		2011 Feb 14 & 287 & 250 \\
		2014 Dec 14 & 195 & 257 \\
		2015 Apr 30 & 180 & 372 \\
		2015 Jul 08 & 180 & 445 \\
		2016 Mar 27 & 180 & 288 \\
		2018 May 10 & 180 & 255\\
		\hline
		\multicolumn{3}{c}{F160W}\\
		\hline
		2011 Jan 20 & 180 & 1110 \\
		2013 Nov 02 & 180 & 510 \\
		2014 Apr 14 & 180 & 372 \\
		2014 Dec 29 & 180 & 361 \\
		2015 Apr 15 & 180 & 400 \\
		2015 Sep 30 & 180 & 478 \\
		2016 Apr 15 & 180 & 390 \\
		2016 Nov 15 & 225 & 304 \\
		2018 May 10 & 180 & 311 \\
		\hline
	\end{tabular}
\end{table}
\fi

\begin{table}
	\centering
	\caption{Definition of epochs for construction of $\Delta$ flats.}
	\label{tab:flats}
	\begin{tabular}{lccr} % four columns, alignment for each
		\hline
		Date & Half-Width  & \#  \\
	         & (days) &  Images \\
		\hline
		\multicolumn{3}{c}{F105W}\\
		\hline
		2011 Feb 14 & 330 & 208 \\
		2014 Dec 14 & 270 & 226 \\
		2015 Apr 30 & 270 & 287 \\
		2015 Jul 08 & 270 & 291 \\
		2016 Mar 27 & 360 & 251 \\
		2018 May 10 & 420 & 203\\
		\hline
		\multicolumn{3}{c}{F160W}\\
		\hline
		2011 Jan 20 & 180 & 984 \\
		2013 Nov 02 & 180 & 308 \\
		2014 Apr 14 & 270 & 300 \\
		2014 Dec 29 & 225 & 247 \\
		2015 Apr 15 & 180 & 305 \\
		2015 Sep 30 & 180 & 403 \\
		2016 Apr 15 & 180 & 252 \\
		2016 Nov 15 & 300 & 258 \\
		2018 May 10 & 180 & 283 \\
		\hline
	\end{tabular}
\end{table}

To generate the $\Delta$ flats, we start with {\tt flt} files, for which the original default flatfield has already been applied. For each image, we generate object masks using the pyraf command {\tt objmasks} with a $3\times3$ convolution kernel \citep{pyraf2012}, masking all 2$\sigma$ detections with an area of $>50$ pixels and growing the mask around each object. We then also apply the persistence masks from \S \ref{sec:persist}. Using these masks, we median scale the {\tt flt} images, median combine them with 5$\sigma$ outlier rejection, and normalize to create output $\Delta$ flats. 
In addition to correcting subtle large-scale trends, the $\Delta$ flats minimize localized "holes" in the response that have developed over time.

\subsubsection{Masking}
The standard astrodrizzle reduction yields a single combined image for each epoch and filter, as well as associated context ({\tt ctx}) and weight ({\tt wht}) images. We first use the {\tt ctx} image, which contains information on which input images contribute to each output science image, to create a mask for each image. We then multiply this mask by the persistence mask.
All images are then transformed to a common reference frame using {\tt scamp} and {\tt swarp} \citep{bertin2006scamp,bertin2010swarp}. Next, all images in a given filter and at a given location (i.e core, parallel, or each bridge field) are  coadded and
SExtractor \citep{bertin1996} is used to construct an object catalog for each location.\footnote{We set the SExtractor detection parameters with a minimum area of 5 pixels and a 2$\sigma$ detection threshold.} These catalogs are used as input to generate elliptical masks which excludes pixels within two Kron radii for all objects in the field, including cluster members, except for the BCG which is left unmasked. 
Additionally, we use a manually generated  region file to mask galaxies near the core of the BCG that are missed by SExtractor. These galaxies, which are identified via visual inspection, are masked with large circular apertures.

Once these masks are generated for each location, they are combined to make a composite mask for the full area in each filter. This composite mask is a union of the individual masks -- any pixel excluded in any of them is excluded in the composite. Finally, the F105W and F160W composite masks are combined to make a final joint mask excluding all pixels masked at either wavelength.

\subsection{Sky Normalization}
At this stage we are ready to correct for relative offsets in the sky level between observations taken at different epochs due to variations in zodiacal light, Earth-shine, and  atmospheric He I emission \citep{brammer2014,pirzkal2014}.  We define a single parallel exposure to be the reference frame, to which we will normalize the sky level for all other exposures. The procedure is straightforward. For each image overlapping with the reference frame, which includes all parallel images and BRIDGE3, we compute the median sky level ($f$) in the unmasked overlap region between that image and the reference image. Each image $i$ is then normalized via
\begin{equation}
f_i^\prime = f_i+ \mathrm{ median}[f_{ref}(\mathrm{overlap})]- \mathrm{median}[f_i(
\mathrm{overlap})].
\end{equation}
Next, we use the same procedure to step through the bridge fields, starting with normalizing the BRIDGE2 images to match the level of BRIDGE3 in the overlap region between the two. We then normalize a single core image  to BRIDGE1, and use this as a reference frame within the core region to which all other core images are normalized. The maximum normalization corrections to the core images to match the parallel reference image were $\sim7-8\times 10^{-3}$ counts s$^{-1}$ for both filters. 

The above procedure results in a robust relative sky calibration from the core to the parallel fields. We note that the results are robust to which parallel image is chosen as the reference image. At this point we only lack an absolute calibration of the sky level. The approach that we take in Section \ref{sec:profiles} is to use data at the largest radii covered by the parallel imaging to define a zero sky level, and verify that the results are robust to the exact choice of annular sky aperture.

\section{Analysis and Results}
\subsection{Radial Profiles}
\label{sec:profiles}
We compute median radial surface brightness profiles in logarithmic bins of width $d\log r=0.15$ starting with an inner radius of 1$^{\prime\prime}$ ($\sim6.6$ kpc).  Use of the median helps to minimize bias arising from local substructures, as pointed out by 
\citet{mansfield2017} and \citet{deason2021}.\footnote{As a test, we also compute the mean profiles and find consistent results.} These profiles are computed individually for each epoch, filter, and field, resulting in a total of 44 separate profiles. We compute the composite profile by calculating the mean and standard deviation for all measurements within each radial bin. The outermost bin in our radial profile is centered at 1.96 Mpc and has an outer radius of 2.31 Mpc. As noted above, one remaining uncertainty is that we have a relative calibration of the sky level in each image, but not an absolute calibration of the sky level. For our 
analysis, 
we compute the mean sky level using data at $r>2.4$ Mpc, making the approximation that the ICL contribution is zero within this region.
It is important to assess the extent to which the minimum radius used to estimate the sky impacts our result; 
varying it  
between 2.4 Mpc and 2.6 Mpc has a negligible impact on the surface brightness profiles or the logarithmic gradients (\S \ref{sec:gradient}). The only difference is that when the minimum radius is set to larger values, the statistical uncertainties in the sky level increase due to the decreased area available for sky determination.

\begin{figure}
    \centering
    \includegraphics[width=9cm]{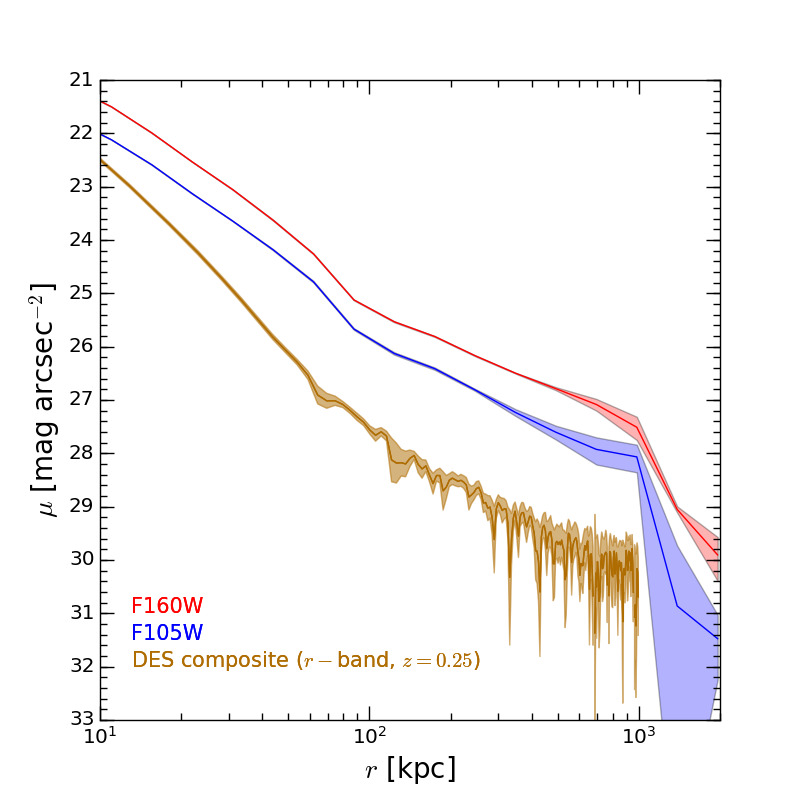}
    \caption{Observed surface brightness profiles for \target\ in the F105W (blue) and F160W (red) filters. The profiles shown are as-observed, with no evolutionary or passband corrections and no correction for cosmological dimming. Data correspond to the median values within logarithmic bins of width $\Delta\log r=0.15$. For comparison, we also include the stacked surface brightness profile from \citet{zhang2019}, which is constructed from $r-$band DES data of $\sim300$ clusters at $z\approx0.25$ (brown). While the normalizations differ due to the differences in redshifts and filters, for both the stacked data and \target\ a transition from the BCG-dominated to ICL-dominated regime can be seen slightly interior to 100 kpc. Both the \target\ and DES data also exhibit a similar profile shape from 100 kpc out to the maximum radii probed by the stacked data.}
    \label{fig:profiles}
\end{figure}

The resultant radial profiles for both filters can be seen in Figure \ref{fig:profiles}. We emphasize that this is the observed profile, with no corrections for cosmological surface brightness dimming, passband ($k-$corrections), or evolution.  Both profiles exhibit a flattening at $r\sim70$ kpc, demarcating the transition from the BCG to the ICL. After this, the profiles show little structure until $r\sim 1$ Mpc, beyond which they decrease rapidly. 

For comparison, we show the stacked Dark Energy Survey surface brightness profile from \citet{zhang2019}, which is an $r-$band profile at $z=0.25$ constructed from an ensemble of $\sim$300 clusters.   The values of the surface brightness are not directly comparable due to the differences in filters and redshift, however the shapes are. The inner transition from BCG to ICL can be seen in both, although the inner slope is somewhat steeper for the composite than for \target. At larger radii, the profiles appear similar out to the largest radii probed by the DES sample. Beyond this, however, we see evidence for a steepening of the profile.

\subsection{Logarithmic Gradient}
\label{sec:gradient}
\begin{figure}
    \centering
    \includegraphics[width=9cm]{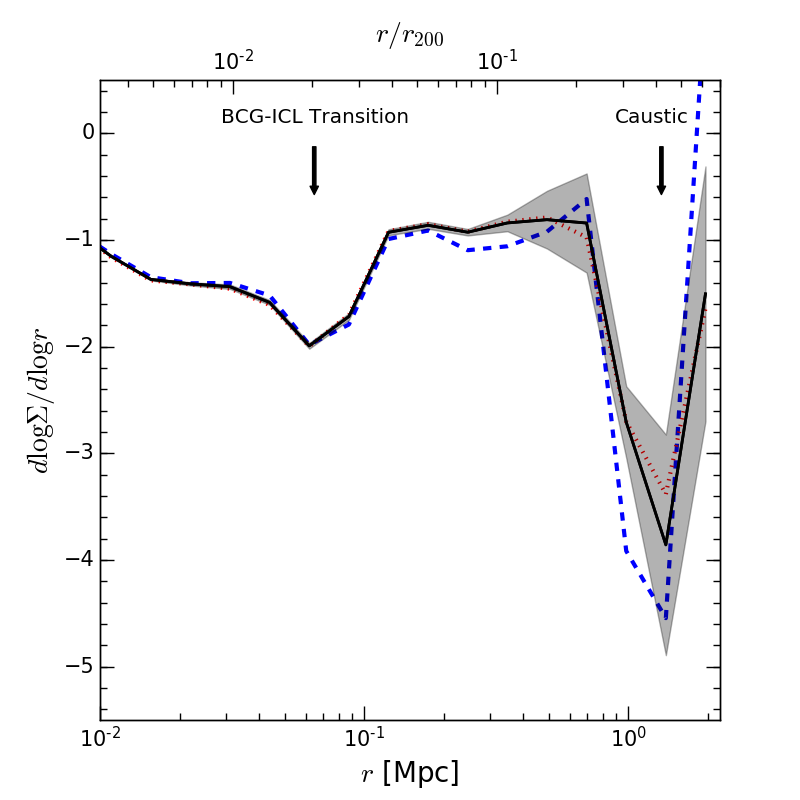}
    \caption{Logarithmic slope of the luminosity surface density. The dashed blue and dotted red curves are for F105W and F160W, respectively. The solid black curve is the uncertainty-weighted mean of the two filters, while the grey region corresponds to the 1$\sigma$ uncertainties.  The transition from the regime dominated by the luminosity of the BCG to that dominated by the ICL corresponds to the inflection at $\sim 60-70$ kpc ($\sim 0.02$ \rtwo). A second, larger dip occurs 
    in the bin centered at 1.4 Mpc (0.44 \rtwo), which spans radii of $1.2-1.7$ Mpc ($0.37-0.52$ \rtwo). This dip is a caustic in the density distribution. Based upon the strength of this dip, we interpret this caustic as plausibly corresponding to the splashback radius. 
    }
    \label{fig:gradient}
\end{figure}

To better quantify the shape, we plot the radial logarithmic gradient of the surface brightness, $d\log\Sigma/d\log r$, 
in both bands in Figure \ref{fig:gradient}, where $\Sigma$ is the flux per square arcsec.  Similar behavior is evident in both filters; we therefore compute a weighted average of the two data sets, which is shown as the solid black curve and associated uncertainties. Two features are clearly evident. First, the transition from the surface brightness being dominated by the BCG to by the ICL can now be clearly associated with the small dip in the gradient at $r\sim70$ kpc. Incidentally, this radius is quite similar to the point at which \citet{montes2021} recently found that the ICL starts to dominate for the nearby cluster Abell 85.  
Second, there is a much stronger feature at $r\sim1.4$ Mpc.  We emphasize that the same feature is seen in both filters, and also verified that a systematic oversubtraction of the sky cannot produce such a feature. On the top axis of Figure \ref{fig:gradient}, we recast the radius in terms of \rtwo, as this is useful for facilitating comparison with theoretical predictions. 
\citet{umetsu2014} derived a weak lensing mass for this cluster of $M_{200m}= 3.4\pm0.7\times10^{15}$ M$_\odot$, which corresponds to \rtwo$=3.18\pm0.22$ Mpc. We use this fiducial value of \rtwo\  for the rest of our analysis.

\begin{figure}
    \centering
    \includegraphics[width=9cm]{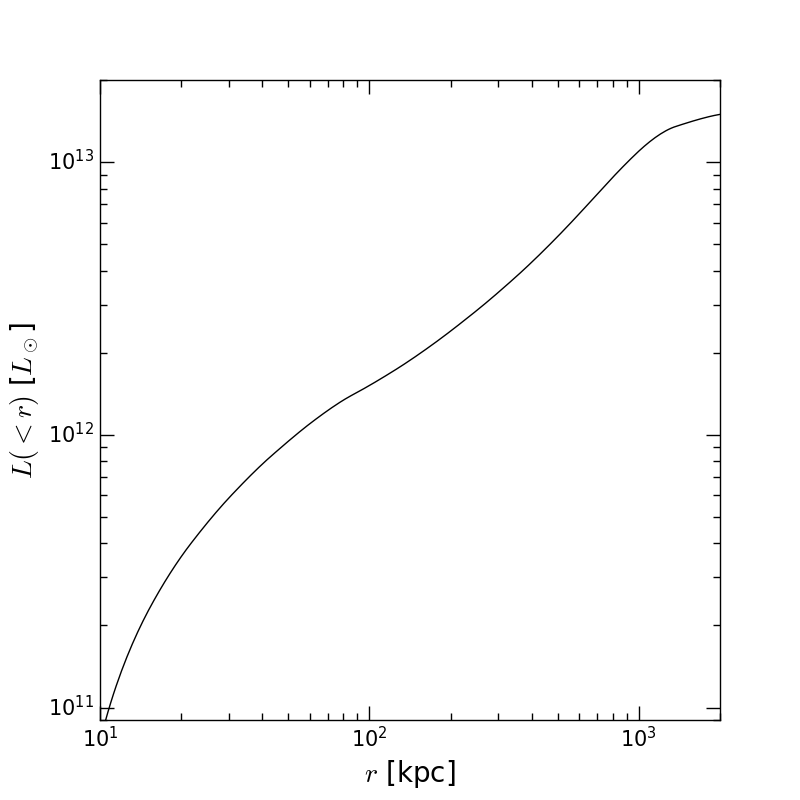}
    \caption{Estimate of the total enclosed luminosity in the BCG+ICL of \target\ as a function of radius. This estimate assumes a radially symmetric surface brightness profile. We omit uncertainties since for a merging cluster like \target\ this systematic uncertainty will dominate the error budget.}
    \label{fig:totlum}
\end{figure}
\begin{figure}
    \centering
    \includegraphics[width=9cm]{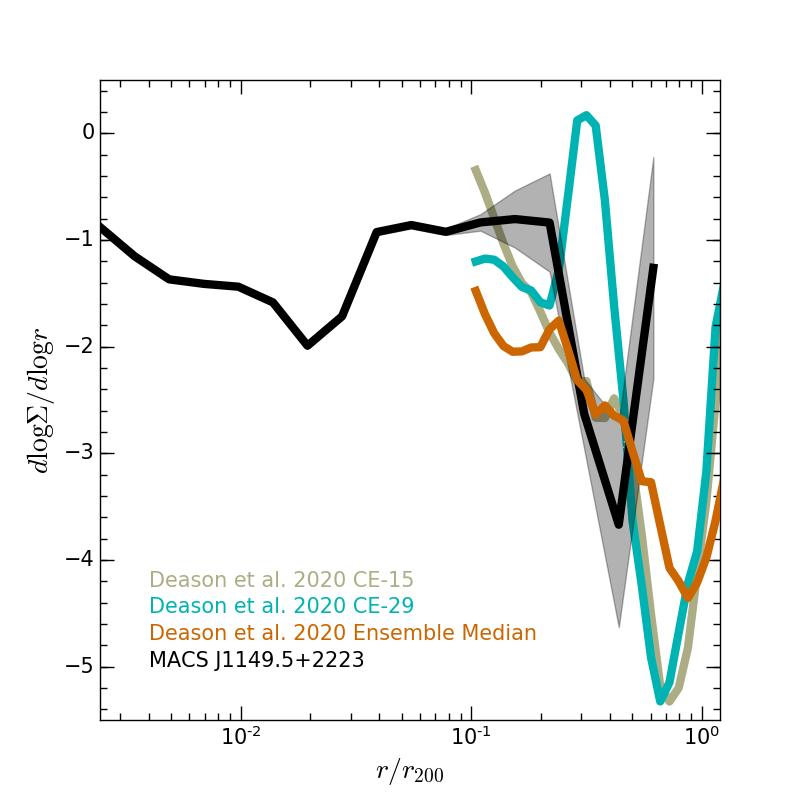}
    \caption{Logarithmic slope of the luminosity surface density as in Figure \ref{fig:gradient}, but now compared with \citet{deason2021} profiles from the C-EAGLE simulation. The three curves shown are all derived using the angular median approach of \citet{mansfield2017}. The curve labeled {\it ``Ensemble Median"} is a stack based upon the full sample of 30 simulated clusters. The other two, CE-15 and CE-19, are individual simulated clusters with high accretion rates. The shape and amplitudes of the caustic feature are similar, but the mimima is at smaller radius for \target\ than for the simulations.
    }
    \label{fig:gradienttheory}
\end{figure}

\subsection{Total Luminosity}
We estimate the total enclosed luminosity as a function of radius, interpolating the F160W surface brightness profiles to bins of $d\log r=0.01$ (Figure \ref{fig:totlum}). A limitation in this calculation is that we lack full angular coverage and therefore make the approximation of a circularly symmetric surface brightness profile. Given the merging state of \target, this is likely to be a poor approximation. Nonetheless, we find that the total luminosity of the BCG+ICL is $\sim1.5\times10^{13} L_\odot$ within 2 Mpc,  indicating that over 90\% (50\%) of the total observed ICL luminosity 
lies outside the central 100 kpc (725 kpc). We also find $L\sim1.3\times10^{13}$ L$_\odot$  
within 1.4 Mpc (\S \ref{sec:gradient} and Figure \ref{fig:gradient}), indicating that there is only a modest contribution to the total luminosity beyond this radius.

\section{Interpretation of the Outer Feature}
Recent theoretical work by \citet{deason2021} predicts the existence of a caustic at the splashback radius ($r_{sp}$), with $r_{sp}\sim r_{200m}$. 
\citet{deason2021} also find secondary caustics at smaller radii for some clusters that experienced significant accretion events in the past.  In Figure \ref{fig:gradienttheory} we compare our observations with the predictions from \citet{deason2021} for the logarithmic gradient in the median surface brightness for an ensemble of clusters at $z=0$ in the C-EAGLE simulations, which span a mass range of M$_{200c}=10^{14}$ to $2.5\times10^{15}$ M$_\odot$, 
The theoretical prediction of the curve for each cluster in the ensemble is calculated by taking the median value for a series of angular wedges within each radial annulus. While slightly different than our median calculation, the methods are sufficiently similar that the profiles should be directly comparable. The depth of the feature observed in \target\ is consistent with those predicted for the splashback radius. The radius of the observed caustic, which lies in the $0.37-0.52$ \rtwo radial bin ($\sim0.44$\rtwo\ bin center),
is smaller 
than would be expected for the splashback radius in a typical cluster. 

There are several possible factors that may explain this difference between the observations and simulations. First, as noted above, secondary caustics do exist for some clusters. We therefore cannot discount the notion that we are viewing a secondary caustic rather than the splashback radius. In the simulated clusters published by \citet{deason2021}; however, the secondary caustic dips are typically very shallow (see their Figures 2-3) and in no instances approach the depth that we see in \target. We therefore consider it unlikely that we are viewing a secondary caustic. 

Second, \target\ is known to be a complex merging system \citep{golovich2016}. 
The ratio $r_{sp}/$\rtwo\ is predicted to be a function of accretion rate, with $r_{sp}/$\rtwo\ decreasing as the accretion rate increases. Because \target\ is a merging system, it is expected that this ratio should be less than unity.
In Figure \ref{fig:gradienttheory} we also include curves for CE-15 and CE-29, which are among the individual simulated clusters with the highest accretion rates. CE-29 is also among the most massive clusters in the simulation ($M_{200m}\sim3.2\times10^{15}$ M$_\odot$), comparable in mass to \target. 
For these clusters the splashback radius moves in as far as $\sim0.7 $\rtwo.  Thus, the dynamical state of the cluster may reduce this discrepancy between the observed splashback radius and simulations. \citet{Diemer2020b} find that at even higher accretion rates $r_{sp}/$\rtwo\ converges to 0.65. Given we observe $r_{sp}\simeq0.37-0.52$ \rtwo, a high accretion rate for \target\ can therefore mostly (but not completely) explain the small observed value of $r_{sp}$.

Third, we must consider \rtwo, which might produce the discrepancy between theory and the observed splashback radius if it is overestimated for our analysis. The \rtwo\ uncertainty may be larger than reflected in the statistical uncertainties due to this merging system's ellipticity (probed here in only one direction) and complex morphology. The ellipticity  
is unlikely to be the explanation. Given that our radial profile is offset by only $\sim27^\circ$ from the major axis of the merger \citep{umetsu2018}, we are more plausibly underestimating than overestimating \rtwo\ along this direction by assuming radial symmetry. Conversely, the complex morphology may provide a partial resolution if \mtwo, and hence \rtwo, has been overestimated along the direction of our observations. The mass within the central 1.5 Mpc is however sufficiently well-determined by multiple groups that this is unlikely to be the full explanation.

The final possibility is that $\Lambda$CDM simulations systematically overestimate cluster splashback radii. For example, self-interacting dark matter would decrease the splashback radius \citep[see][for a discussion of this topic]{more2016}.  
Published measurements based upon galaxy density and luminosity profiles for various cluster samples however yield ensemble values in the range of $r_{sp}/$\rtwo$\simeq0.8-1.2$ \citep{more2016,chang2018,shin2019,murata2020,bianconi2020} -- all significantly larger than what is observed in this instance.

\section{Conclusions}
This paper is the first in a series investigating the properties of the intracluster light in the \hubble\ Frontier Field cluster \target. We have used the combination of existing data sets and new GO observations to provide contiguous radial coverage from the cluster center to 2.8 Mpc, which corresponds to $\sim0.88$ \rtwo, in a strip extending southeast from the cluster core. The focus of this paper is the radial surface brightness profile for the cluster, and we also present the methodology employed in our analysis.  
Our central findings are: 
\begin{itemize}
    \item We are able to extract the radial surface brightness profile of the brightest cluster galaxy and intracluster light out to a radius of 2 Mpc, approximately 0.6 \rtwo. This distance is the furthest to which the BCG+ICL profile has been measured for any individual cluster, and is comparable to the distance probed by the stacking analysis of \citet{sampiaosantos2020} for 528 DES clusters. The key to reaching these large radii for \target\ lies in bridging the gap between the HFF primary and parallel fields to enable a robust sky determination. 
    \item We see a distinct transition in the profile at $\sim70$ kpc, which we interpret as the radius at which the profile transitions from the BCG to the ICL.
    \item We identify a sharp steepening of the ICL surface brightness profile beyond 1 Mpc. From the logarithmic derivative of the luminosity density, we find evidence that this transition corresponds to a caustic in the stellar distribution at $\sim1.2-1.7$ Mpc (approximately $0.37-0.52$ \rtwo). The center of the radial bin in which the caustic is found is 1.4 Mpc (0.44 \rtwo).
    \item Under the assumption of a radially symmetric profile, we calculate that the total luminosity of the BCG+ICL is
    $\sim1.5\times10^{13} L_\odot$ within 2 Mpc. 
    The luminosity profile indicates that over $90$\% ($50$\%) of the total BCG+ICL luminosity lies outside of the central 100 (725) kpc. 
    \item Recent work by \citet{deason2021} has argued for the existence of caustics in the ICL at the cluster splashback radius.
    While those simulations predict that the splashback radius should typically be at $\ga 0.7$\rtwo, the strength of the feature in the logarithmic derivative plot for \target\ is consistent with expectations for the splashback signature. 
    If the observed caustic is indeed the splashback radius, the offset relative to the simulations may arise in part from \target's complex dynamics -- the splashback radius in a merging system should be smaller relative to \rtwo.
\end{itemize}

While this work is the first to detect the ICL beyond 1 Mpc for an individual cluster, similar studies are now possible with minimal additional required data for the remaining HFF clusters. 
Detections of similar caustics should be achievable in at least some of these systems, and will  further shed light on
whether these features correspond to splashback radii.  Looking forward, 
upcoming next-generation facilities including \euclid\ and the {\it Roman Space Telescope} 
have the potential to enable systematic studies of the extended intracluster light distribution for large numbers of clusters and galaxy groups out to the splashback radius.  

In the shorter term, this paper is the first on the ICL in \target. The partitioning of baryons between the ICL, galaxies, and gas, and the distribution of ICL relative to dark matter will be investigated in subsequent work.

\section*{Acknowledgements}
 This work is based on observations with the NASA/ESA {\it Hubble Space Telescope} obtained at the Space Telescope Science Institute, which is operated by the Association of Universities for Research in Astronomy, Incorporated, under NASA contract NAS5-26555. Support for Program number 15308 was provided by NASA through a grant from the Space Telescope Science Institute, which is operated by the Association of Universities for Research in Astronomy, Incorporated, under NASA contract NAS5-26555. The work of T.C. was carried out at the Jet Propulsion Laboratory, California Institute of Technology, under a contract with NASA.

%%%%%%%%%%%%%%%%%%%%%%%%%%%%%%%%%%%%%%%%%%%%%%%%%%

\bibliographystyle{mnras}
\bibliography{ms} % if your bibtex file is called example.bib

% Don't change these lines
\bsp	% typesetting comment
\label{lastpage}
\end{document}